# The Link between Magnetic Fields and Filamentary Clouds: Bimodal Cloud Orientations in the Gould Belt

Hua-bai Li[1, 2], Min Fang[3], Thomas Henning[1] & Jouni Kainulainen[1]
1. Max-Planck-Institut für Astronomie, Königstuhl 17, 69117 Heidelberg, Germany
2. Department of Physics, The Chinese University of Hong Kong, Hong Kong; hbli@phy.cuhk.edu.hk
3. Purple Mountain Observatory, 2 west Beijing Road, Nanjing, 210008, China

**Abstract**
The orientations of filamentary molecular clouds in the Gould Belt and their local ICM (inter-cloud media) magnetic fields are studied using near-infrared dust extinction maps and optical stellar polarimetry data. These filamentary clouds are a few-to-ten parsecs in length, and we find that their orientations tend to be either parallel or perpendicular to the mean field directions of the local ICM. This bimodal distribution is not found in cloud simulations with super-Alfvénic turbulence, in which the cloud orientations should be random. ICM magnetic fields that are dynamically important compared to inertial-range turbulence and self-gravity can readily explain both field-filament configurations. Previous studies commonly recognize that strong magnetic fields can guide gravitational contraction and result in filaments perpendicular to them, but few discuss the fact that magnetic fields can also channel sub-Alfvénic turbulence to form filaments aligned with them. This strong-field scenario of cloud formation is also consistent with the constant field strength observed from ICM to clouds (Crutcher et al. 2010) and is possible to explain the "hub-filament" cloud structure (Myers 2009) and the density threshold of cloud gravitational contraction (Kainulainen et al. 2009).

## 1. Introduction

Self-gravity, turbulence, and magnetic fields (B-fields) all play a role in the star-formation process (e.g., McKee & Ostriker 2007), which transforms only a small fraction of the mass of molecular clouds into stars (e.g., Lada et al. 2010). How exactly these forces interact with each other to regulate the conversion from gas to stars is still far from clear. Recent observational studies (e.g., Zhang et al. 2009; Schmalzl et al. 2010; Wang et al. 2012), and especially those exploiting the Herschel satellite data (e.g., André et al. 2010; Molinari et al. 2010; Henning et al. 2010; Hill et al. 2011; Ragan et al. 2012), have supported the decades old *"bead string"* scenario of star formation (e.g., Schneider &

Elmegreen 1979; Mizuno et al. 1995; Nagahama et al. 1998). In this picture, molecular clouds first form filaments (parsec- to tens-of-parsecs long *"strings"*; Fig. 1 & 2), which then further fragment into dense cores (*"beads"*). This scenario emphasizes the significance of filamentary structures as a critical step in star formation.

However, the formation mechanism of filamentary clouds is still not understood. One model for filament formation is shock compression due to stellar feedback, supernovae, or turbulence (e.g., Padoan et al. 2001, Hartmann et al. 2001; Arzoumanian et al. 2011). While shock compression might be an efficient mechanism for mass collection, this model is in contradiction with a morphological observational constraint for filamentary cloud formation: molecular clouds commonly show long filaments *parallel* with each other (Myers 2009). Stellar feedback (see, e.g., Figure 3 of Hartmann et al. 2001) and isotropic super-Alfvénic turbulence (see, e.g., Figure 2 of Padoan et al. 2001) cannot explain these parallel cloud filaments. Also, the streaming motions predicted by the shocked cloud model are not generally observed (e.g., Loren 1989). While filamentary clouds in super-Alfvénic simulations will align with the B fields *within* them due to shock compression, there is no correlation between the directions of clouds and mean B-field directions from their surrounding ICM (see Figure 2 of Padoan et al. 2001), a property that we will test in this paper.

There are two other possible mechanisms to form filamentary clouds, which both require dynamically important B fields. These are *B-field channeled* gravitational contraction (e.g. Nakamura & Li 2008) and anisotropic sub-Alfvénic turbulence (e.g. Stone, Ostriker & Gammie 1998). In the former mechanism, the Lorentz force causes gas to *contract* significantly more in the direction along the field lines than perpendicular to the lines, if the gas pressure is not strong enough to support the cloud against self-gravity along the B field (Mouschovias 1976). This contraction will result in flattened structures, which look elongated on the sky (see, e.g., Nakamura & Li 2008). When there are multiple contraction centers, the gas will end up in parallel filaments. Sub-Alfvénic anisotropic turbulence has the opposite effect: turbulent pressure tends to *extend* the gas distribution more in the direction along the field lines, and leads to filaments aligned with the B field (see Figure 2 of Stone, Ostriker & Gammie 1998; Cho & Vishniac 2000; Cho & Lanzarian 2002; Vestuto et al. 2003; Li et al. 2008). This means that the *competition between gravitational and turbulent pressures in a medium dominated by B fields* will shape the cloud to be *elongated either parallel or perpendicular to the B fields* when reaching the equilibrium (see Fig. 3 for a schematic summary of this scenario). Different from the shocked cloud model, streaming motions are not expected in the equilibrium stage, which agrees with observations from, e.g., Loren (1989). A similar picture can result from gravitational instability of a thin gas layer, which is pressure-confined, *non-turbulent*, isothermal, and, most importantly, with *uniform* (i.e., dynamically dominant) B fields that are parallel with the gas layer (Nagai et al. 1998).

Recently, the presence of ordered dynamically important B fields required in the scenarios discussed above got significant observational support. First, it has been shown (e.g., Han & Zhang 2007; Reid & Silverstein 1990) that the direction of the line-of-sight component of the Galactic B field is preserved in molecular clouds, using the correlation between the Zeeman splitting data of masers and rotation measures of pulsars. Second, Li et al. (2009) found that the plane-of-sky components of Galactic B-field directions are also preserved in cloud cores, based on the correlation between the polarization directions of sub-mm data from cloud cores (sub-pc scale) and optical data which probes the ambient ICM (hundred-parsec scale). The picture that galactic B fields anchor into molecular clouds is also supported by a study of the face-on galaxy M33 (Li & Henning 2011) using the polarization of CO emission lines. Finally, Zeeman measurements (Crutcher et al. 2010) show that the B-field strength is quite constant from the ICM to lower-density regions of molecular clouds ($N_H \sim 5\times10^{21}$ cm$^{-2}$; see more discussion in section 4.2). These observations imply that molecular clouds are threaded by ordered B fields, which are not tangled by self-gravity or turbulence during the cloud and core formation processes and are not compressed in regions with $N_H < 5\times10^{21}$ cm$^{-2}$, which comprise most of the volume of a molecular cloud (Kainulainen et al. 2009; Lada et al. 2010). This ordered B field should, in turn, channel the turbulent and gravitational gas motion, such that the resulting cloud shapes are elongated in directions either parallel or perpendicular to the local ICM B fields.

This work aims to test the scenario of B-field channeled formation of filamentary molecular clouds (Fig. 3) using as a diagnostic the relative alignments between filamentary morphologies of clouds in the Gould Belt and their nearby ICM B-field directions. In particular, we concentrate on the large-scale (parsecs to tens-of-parsecs) filamentary structures that are characterized by relatively low column densities ($A_v < 10$ mag, $\langle A_v \rangle \approx 2$ mag; e.g., Kainulainen et al. 2009). Therefore, we specifically want to analyze B-field direction in the low density ICM, not in the dense structures nested inside molecular clouds. We probe filamentary cloud structures with dust extinction maps (Dobashi 2011) and their ambient *ICM B-field* directions using optical stellar polarimetry data (Heiles 2000).

In section 2, we describe how the targets are selected and how the filament/field directions are measured. The resulting directions are compared in section 3 to test the filament formation scenario. In section 4, we discuss our findings in the context of other recent observational studies, including the star formation threshold, cloud contraction threshold and the "hub-filament" structures (Myers 2009). A summary is given in section 5.

## 2. Observation and data analysis

We chose the molecular clouds in the Gould Belt for this study. They are nearby (150-500 pc; Loinard et al. 2011), thus providing the highest possible spatial resolution, and most of them are located at high Galactic latitudes ($|b| > 10^\circ$). Since stellar polarization samples the B-fields along the entire line-of-sight weighted by dust density, it is important to disentangle the molecular clouds in question from physically unrelated material along the line-of-sight. At high Galactic latitudes, the contribution to the total polarization from dust that is not related to the target cloud is relatively small (see section 4.1 for more discussion about optical polarimetry) simply because there is less dust at higher latitudes. Our sample clouds are listed in Table 1. We attempted to study all the 13 clouds[1] involved in the Herschel Gould Belt Survey (André et al. 2010), but there is no data available for the B fields of Polaris Flare.

As we want to test the cloud formation scenarios by comparing the orientations of clouds and ambient ICM B fields, in the following we describe how these orientations are determined.

---

[1] http://starformation-herschel.iap.fr/gouldbelt/

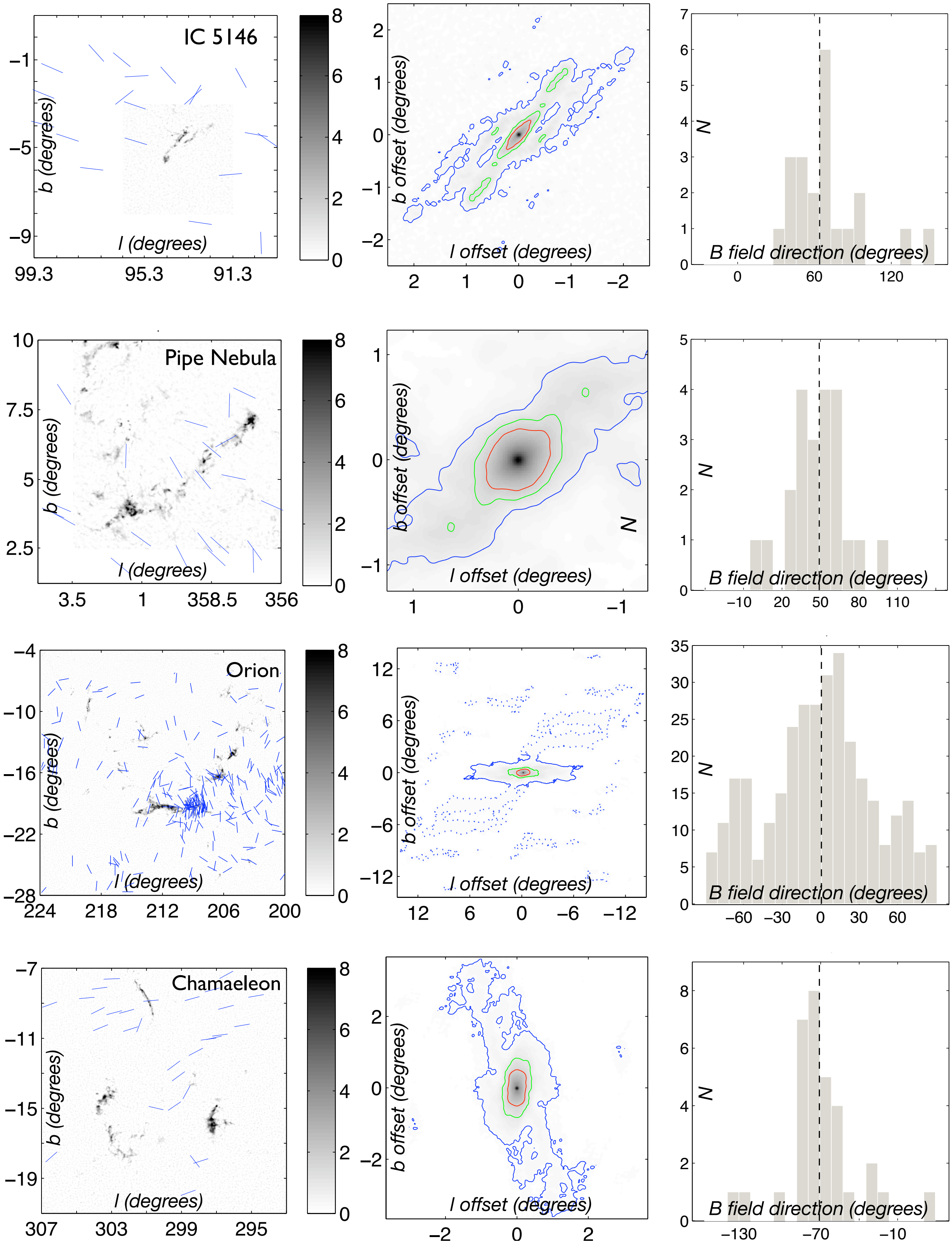

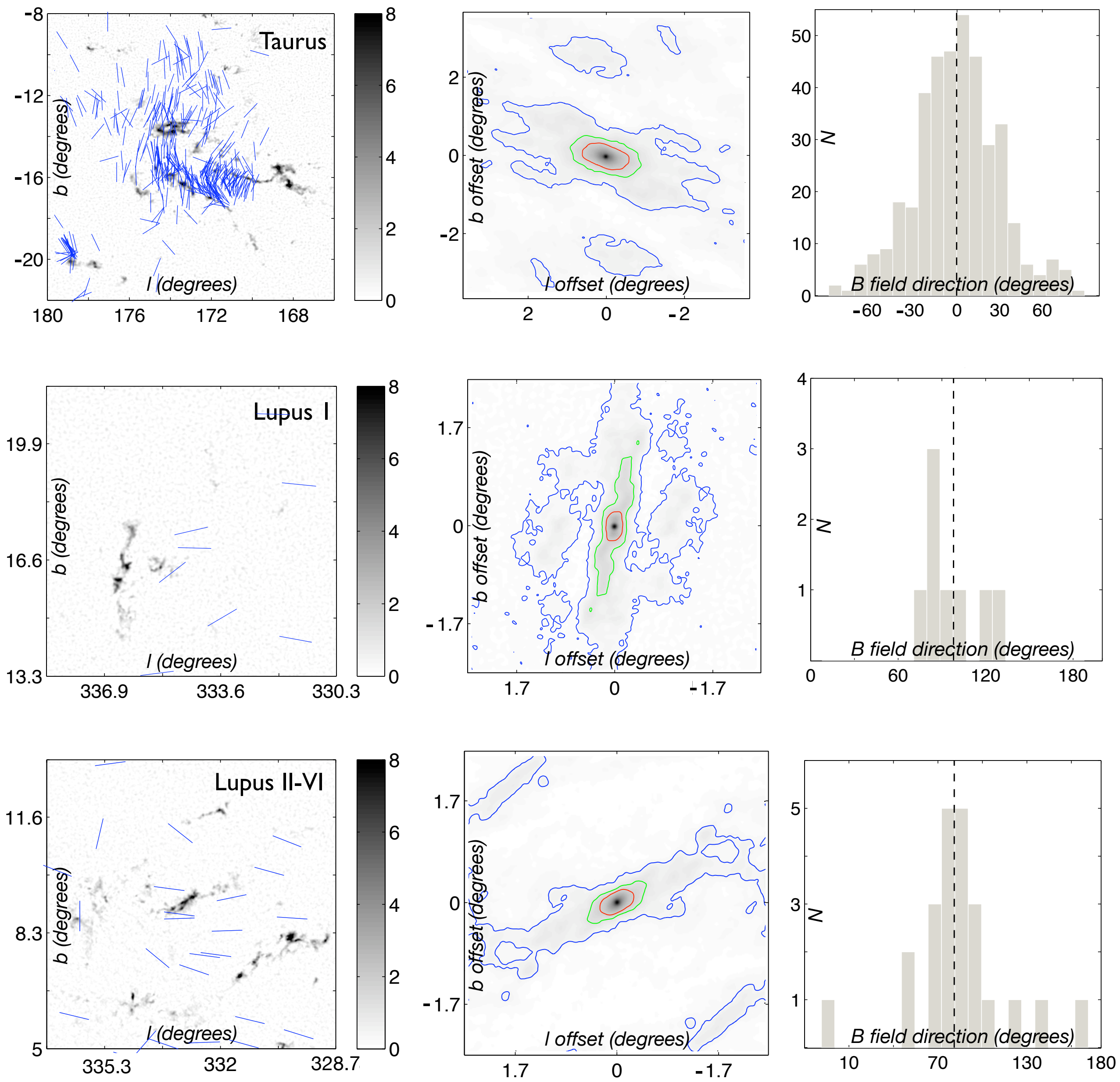

**Fig. 1** *Left column*: $A_V$ maps from Dobashi (2011) in Galactic coordinates (degrees) overlapped with optical polarization (blue vectors) from Heiles (2000). Note that the sensitivity of the Heiles archive mainly traces $A_V < 2$ mag (from ICM to low-density parts of a cloud), so the positions of the vectors (which might overlap with regions of $A_V > 2$) cannot be an indication of the column densities that have been probed by optical polarimetry. Many of these 6 clouds (7 regions) consist of parallel filaments, indicated by the parallel blue contours in the middle column. *Middle column*: The autocorrelations of the $A_V$ maps. The coordinates are the offsets in degrees. The three contours are, respectively, of value 1 (blue), 1/5 peak value (red), and the value in between (green). The linear fit to the region within a contour is used to define the direction of the cloud. Most cloud directions do not vary much with the density. For Orion, there are clearly several blue parallel contours, and we fit to the central one (the solid contour) only, so the relative positions of the filaments will not affect the direction fit. *Right column*: Distributions of the B-field orientations inferred from the optical polarimetry data shown

in the left column, measured counter-clockwisely from the Galactic North in degrees. A dashed line shows the Stokes mean of all the field detections in a map (section 2.2), i.e., the mean field direction. Most cloud directions are nearly perpendicular to the mean field directions.

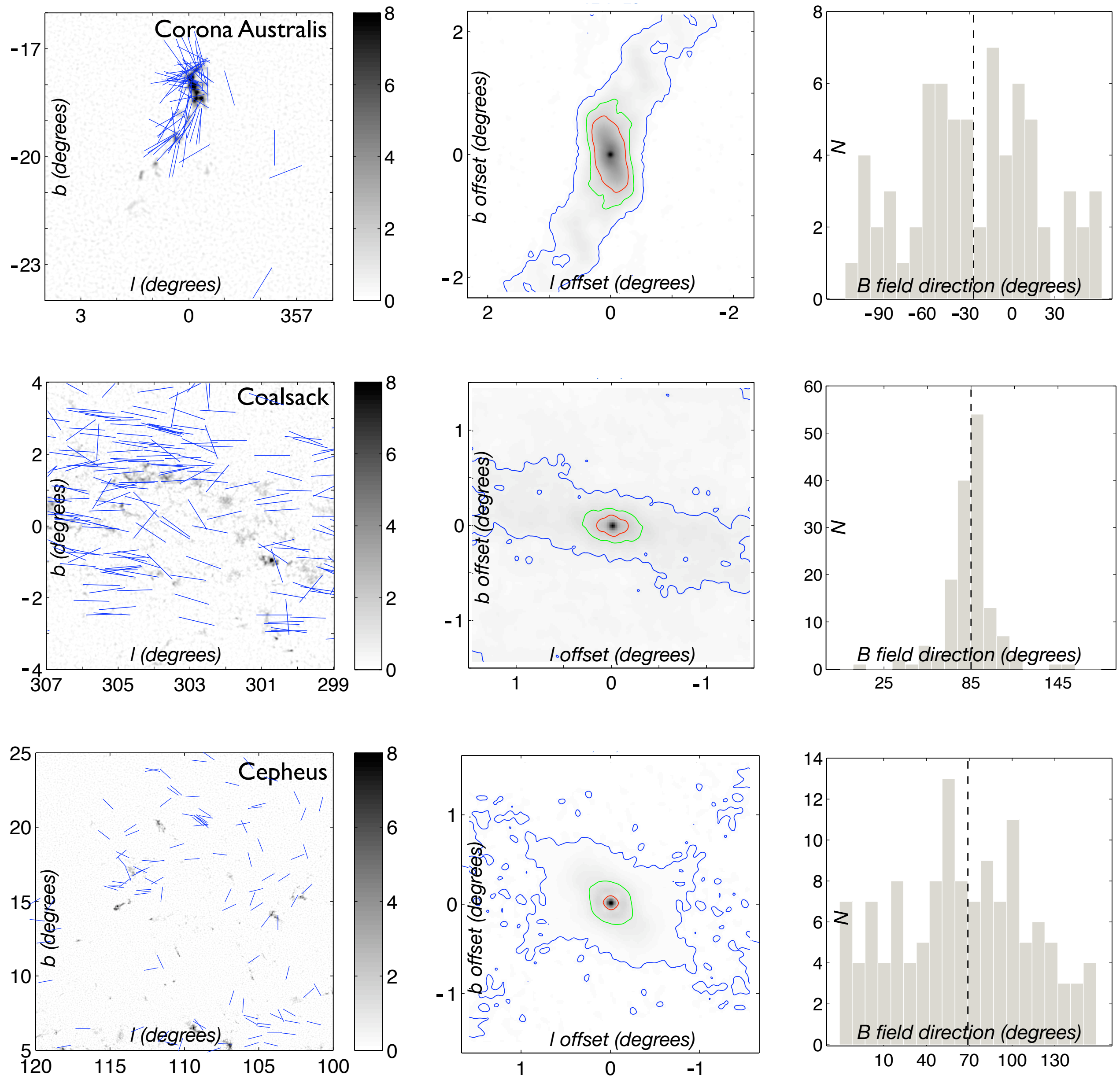

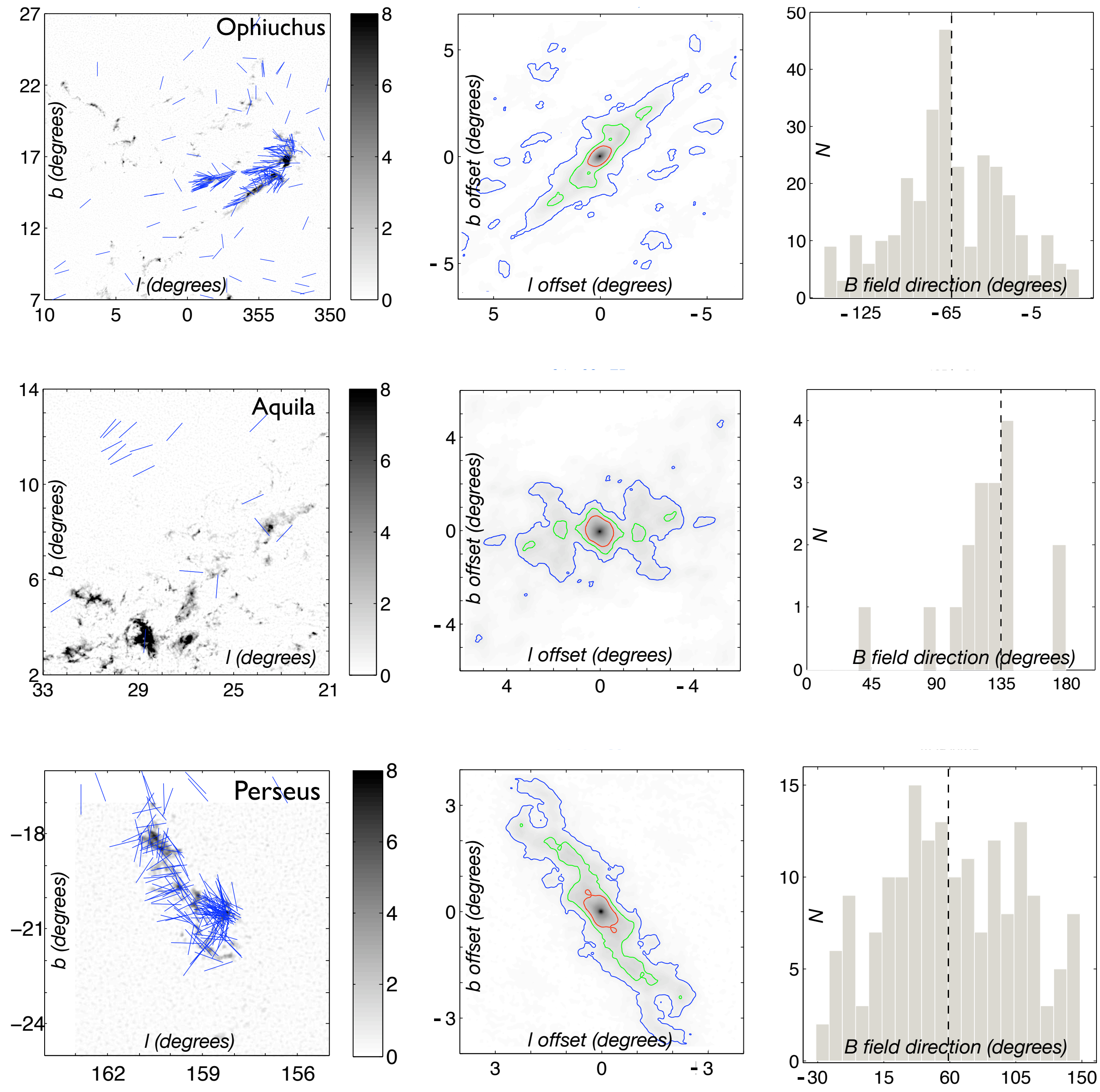

**Fig. 2** Similar to Fig. 1, but these 6 clouds are mostly with directions aligned with the mean field directions. Aquila is a special case, where the cloud direction changes approximately 90° from low to high density (see the discussion in section 4.3).

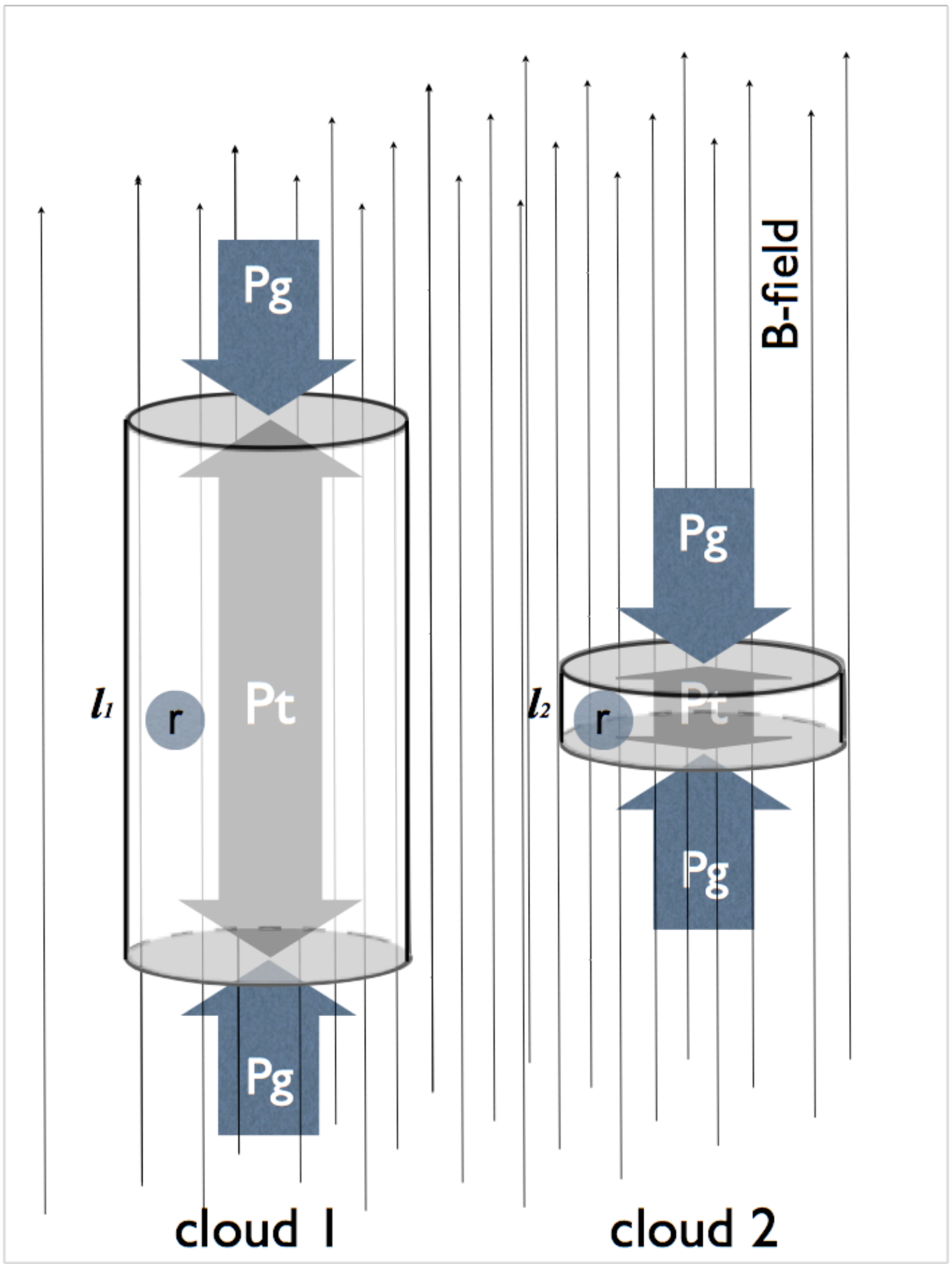

**Fig. 3 A schematic illustration of the two types of filamentary clouds forming under dynamically dominant B fields.** By dynamically dominant, we mean that gravity is magnetically subcritical and the turbulence is sub-Alfvénic. Gravitational pressure ($P_g$) and turbulent pressure ($P_t$) here only effectively compete along the direction of the field lines. When reaching equilibrium ($P_g = P_t$), the dimension of the clouds along the B field are respectively $l_1$ and $l_2$. Projected on the plane of sky, cloud 1 is parallel with the B field and cloud 2 is perpendicular to the B field. A subregion (with diameter r) within these clouds has to reach a critical density in order to contract in a direction across the field lines (see section 4.2).

**Table 1** Directions in degrees (increasing counterclockwisely from Galactic North) of filaments and B fields

| Cloud Name | Filament | | | B fields | | |
|---|---|---|---|---|---|---|
| | high[a] | medium[a] | low[a] | mean[b] | IQR[b] | number of detections |
| IC 5146 | -50 (3.8) | -53 (3.7) | -38 (5.1) | 64 | 32 | 21 |
| Pipe Nebula | -41 (1.4) | -44 (1.6) | -45 (3.4) | 49 | 27 | 22 |
| Orion | -85 (2.0) | -91 (2.8) | -90 (4.3) | 1 | 54 | 325 |
| Chamaeleon | -184 (1.9) | -187 (2.1) | -161 (10.5) | -71 | 22 | 32 |
| Taurus | -105 (2.1) | -105 (2.3) | -105 (4.1) | 0 | 37 | 387 |
| Lupus I | -11 (1.7) | -14 (5.4) | -1 (6.9) | 98 | 26 | 8 |
| Lupus II-VI | -63 (2.0) | -60 (2.6) | -73 (7.1) | 81 | 23 | 23 |
| Corona Aus. | 12 (2.4) | 1 (2.6) | -26 (4.6) | -27 | 65 | 69 |
| Coalsack | 86 (1.6) | 80 (1.8) | 74 (4.3) | 85 | 13 | 222 |
| Cepheus | (1) | 45 (1.3) | 65 (1.6) | 69 | 73 | 128 |
| Ophiuchus | -38 (2.0) | -48 (3.6) | -45 (4.6) | -65 | 51 | 305 |
| Aquila | 26 (1.3) | 92 (1.3) | 105 (2.7) | 135 | 21 | 17 |
| Perseus | 38 (2.1) | 31 (6.5) | 32 (7.2) | 59 | 69 | 170 |

**a.** High-, medium- and low-density directions are defined by, respectively, the red, green and blue autocorrelation contours in the middle column of Fig. 1 & 2. (section 2.1). The numbers in parentheses are the aspect ratios of the filaments.

**b.** Based on optical polarimery data shown in the right column of Fig. 1 & 2 (section 2.2)

## 2.1 Orientations of the filamentary clouds

*The definition of cloud orientation*

Determining the orientation of a molecular cloud is not a straightforward task; the result undoubtedly depends somewhat on the tracer and how the filaments are defined. Intuitively, along the direction of cloud elongation, the correlation length of the cloud map should have a maximum. We thus define the orientation of a cloud by the long-axis direction of the autocorrelation function of the column density map.

In practice, we first Fourier transform the extinction map, and multiply the resulting transform by its complex conjugate. The inverse transform of the product gives the autocorrelation function, which is then normalized by the product of the square of the mean column density and the pixel number. For a given *contour* of this normalized autocorrelation map, the long-axis direction can be obtained by fitting a linear function to the pixel positions located within the contour, such that the summation of all the distances between the linear function and the pixels is minimized. Using various values of *contours*, we can examine the dependence between cloud directions and column densities.

*The extinction maps*

With the definition of a cloud orientation given above, we have derived cloud directions using dust extinction maps from three archives: the IRAS-based extinction maps from Schlegel, Finkbeiner & Davis (SFD; 1998) and near-infrared color-excess 2MASS-based extinction maps from Rowles & Froebrich (RF; 2009) and from Dobashi (2011). The difference between the latter two archives is that Dobashi applied a 2˚ high-pass filter to the maps in order to remove the extended, low-column density component that clearly results from the diffuse extinction of the Galactic plane instead of clouds.

The three archives have similar angular resolutions (a few arcmin), and result in similar directions for clouds with galactic latitude $|b| > 10˚$. For the 5 clouds with $|b| < 10˚$ (Pipe, IC 5146, Lupus II-VI, Coalsack and Aquila; Fig. 1 & 2), the SFD and RF cloud maps show significant extended fore-/background contamination from the Galactic disc, and the directions derived from their autocorrelation trace the Galactic plane instead of cloud structures. The fact that the Dobashi maps result in similar orientations as those derived from SFD and RF maps for clouds with $|b| > 10˚$, but not $|b| < 10˚$, suggests that high-pass filtering efficiently removes only the extended component.

It could be argued that Dobashi's 2˚ spatial filter may affect the resulting directions for clouds with $|b| < 10˚$. To look into this, we applied 1˚ and 4˚ high-pass filters to the RF maps and re-calculated the orientation angles. The resulting angles are indistinguishable from those derived from the original Dobashi (2011) data. We thus adopt the Dobashi data for the analysis of cloud orientation.

*Results*

The Dobashi maps of the Gould Belt clouds and their normalized autocorrelation maps are presented in Fig. 1 & 2. We show in each autocorrelation map three contours: unity, 20% peak value, and the mean of these two values. The orientations defined by these contours are listed in Table I. Above the 20% peak values, some clouds (e.g., Cepheus) become roundish. The differences between the orientations derived from different contours within a cloud are less than 15˚ for most (9) cases. Since gas may rotate during contraction (Hartmann & Burkert 2007), which may contribute partly to the direction differences, the errors in our direction measurements must be smaller than 15˚; see section 4.3 for more discussion on the direction variation from diffuse to dense regions.

Finally, we note that Coalsack is composed of two clouds along the line of sight at very different distances (Beuther et al. 2011). So to some degree, the elongation is due to the arrangement of the two clouds on the sky.

## 2.2 Orientations of the ICM B fields

It is empirically found that the polarization direction of optical starlight is aligned with the B-field orientation in the ICM with $A_V < 3$ mag (e.g., Arce et al. 1998, Poidevin & Bastien 2006). We use the optical polarimetry catalog published by Heiles (2000), which is the most comprehensive database to date. The catalog contains ~1700 detections that fall in the areas considered in this work. For these detections, ~40% come with measurements of E(B-V). Adopting Av/E(B-V)=3.1 to convert the reddening to extinction, we can estimate that about 85% of the detections are with Av below 2 mag. This means that the polarimetry detections of Heiles indeed probe mainly the ICM.

As our aim is to compare orientations of molecular clouds to their *ICM B-field* directions, the Heiles data forms an excellent basis for our study. While the B field directions *inside* the cloud filaments are *not* needed for our investigation, we note that Li et al. (2009) showed that ICM B-field directions anchor into clouds all the way down to cores (see more discussion in section 4.1)

To calculate the mean field direction of the local ICM, we use the "equal weight Stokes mean" (Li et al. 2006) of all the polarization detected in the map. The method is summarized in the following:

1. For each detected polarization direction $\theta_i$ (increasing counterclockwisely from the Galactic North) within a cloud map (Fig. 1 & 2, right column), the Stokes parameters are defined as $q_i = \cos(2\theta_i)$ and $u_i = \sin(2\theta_i)$. The polarization fraction of the detection is ignored, because we do not want to weigh a sight line more simply because its grain alignment efficiency is higher.
2. The mean direction is then calculated from the Stokes parameters $\sum q_i$ and $\sum u_i$.

The resulting mean B-field directions are listed in Table I, together with the interquartile ranges (IQRs, the difference between the upper and lower quartiles) of the detections in each region.

The most rigorous way to derive the ambient ICM B-field direction for a cloud using optical polarimetry is to subtract the polarization of the local foreground stars from that of the local background stars (e.g., Li et al. 2006). However this is not practicable for most clouds in our sample, because there are not many local stars in the archive for the clouds. However, most clouds in our sample are at high Galactic latitudes and diffuse dust component toward them is very thin. Consequently, the foreground contamination to the mean polarization should not be critical. Also, any contamination would only increase the uncertainty of B-field orientation measurement, but would not introduce an artificial correlation with the clouds.

Stellar feedback, presenting to some degree in most our clouds, can quite possibly change B-field orientations and increase the dispersions (IQRs). This might explain the larger B-field IQRs in regions with higher star formation activities, e.g., Perseus (Bally et al. 2008). Since this is a potential randomizing process of B fields, the pre-stellar field-filament correlation can be even stronger, in case we find any with the current data sets.

After Heiles (2000), more polarimetry data has been collected for individual clouds (e.g., Alves et al. 2008; Pereyra & Magalhaes 2004), and they all agree with the mean B field directions shown in Figs. 1 and 2. This is an indication that the Heiles data is representative, which is not a surprise given that its typical field dispersion is only ~ 30° (see section 4) within regions so extended (Figs. 1 and 2) that new data is collected from their subregions.

**3. Results**

The cloud and B-field directions derived in the previous sections are compared in Fig. 4. The horizontal error bars stand for the IQRs of the field directions. The vertical arrows show the range of cloud orientations listed in Table 1: the tails correspond to directions of almost the entire maps (regions with autocorrelation > 1), and the heads point at higher densities (regions with autocorrelation > 20% peak values). Fig. 4 also shows the relations y = x and y = x-90°, i.e., parallel and perpendicular alignments. The shaded regions along these relations show the average (42˚) of the B-field IQRs from all clouds. All pairs of mean fields and cloud directions (tails) fall within 30° from being either parallel or perpendicular to each other.

We performed Monte Carlo simulations to study how significant this result is different from independent (random) cloud and B field orientations. Randomly picking 13 pairs of vectors (standing for our 13 pairs of cloud and B field directions) $10^6$ times, projecting the vectors to a plane, we studied how many pairs out of the 13 have an offset within 30° from either parallelism or perpendicularity. It happens that only 0.6% of the $10^6$ simulations have all 13 offsets within 30°. This is an indication that random orientation is unlikely and the clouds are preferentially aligned either perpendicularly to or parallel with the B field directions.

If the clouds are not randomly oriented, we should further study what is the typical offset between cloud and B-field directions. Again we use Monte Carlo simulations, assuming that cloud orientations follow a double-gaussian distribution which peaks at the directions perpendicular and parallel with the B fields. For a given standard deviation ($\sigma_i$) of the gaussian distribution, 13 values are randomly selected from the distribution to simulate the cloud-field offsets. Vector pairs holding each of the 13 offsets are randomly picked and projected on a plane to simulate 2-D offsets. $10^4$ simulations have been carried out to estimate $P(obs \mid \sigma_i)$, the probability to obtain the observed condition (all 13 2-D offsets are within 30° from either 0° or 90°) for a given $\sigma_i$. Then, based on Bayes′ theorem, the probability of $\sigma_i$ given our observation is

$$P(\sigma_i \mid obs) = P(obs \mid \sigma_i)P(\sigma_i)/P(obs) = P(obs \mid \sigma_i)\Big/\sum_{i=1}^{N} P(obs \mid \sigma_i).$$

$P(\sigma_i \mid obs)$ of $\sigma_i$ between 1° and 37° are shown in Fig. 5, along with the cumulative probabilities. The 95 % confidential range for the STD of the 3-D offset from either parallelism or perpendicularity is around 20°.

## 4. Discussion

The observed bimodal distribution of the cloud and B-field alignments can be used as a diagnostic between different filament formation scenarios. If the primary driver of filament formation is shock compression from super-Alfvénic turbulence, i.e., if B-fields are dynamically unimportant, there should be no alignment between the clouds and ICM B fields. This is because super-Alfvénic turbulence can compress gas in any direction and form filaments regardless of the large-scale ICM B fields. *The bimodal correlation is clearly in disagreement with such a situation, and rather supports a picture in which the B-fields are dynamically important* (see Fig. 3).

The B-field direction IQR is also informative. It takes only a slightly super-Alfvénic turbulence to make the B-field morphology random in numerical simulations. The transition from ordered to random field morphologies is quite sensitive to the Alfvénic Mach number ($M_A$, the ratio of turbulent to Alfvénic velocity). For example, Falceta-Gonçalves et al. (2008) showed that B-field morphologies are ordered for $M_A = 0.7$ but random for $M_A = 2$ (no value in between was shown). Random directions will have an IQR ~ 90°, which is much larger than the mean IQR = 42° we observed.

Following Chandrasekhar and Fermi (1953), we can also estimate the *lower limit* of B-field direction dispersions with super-Alfvénic turbulence. Assuming that the STD of B-field direction (σ, observed with a line of sight perpendicular to the mean field) is *completely* due to gas turbulence, Chandrasekhar and Fermi derived the relation (CF relation) : $\sigma$ (radians) = $[4\pi\rho]^{1/2}\,\upsilon/B$, where B, ρ and υ are, respectively, the B-field strength (Gauss), density ($gm/cm^3$) and line-of-sight turbulent velocity (cm/s). They used small angle approximation, and numerical simulations (e.g., Ostriker et al. 2001) indicated that the CF relation is a good approximation only when $\sigma < 25°$. Assuming that the B-field directions follow a gaussian distribution, IQR = 42° from our sample ICM is equivalent to *STD = 31°*, which is too large for using the CF relation. Falceta-Gonçalves et al. (2008) improved the relation by substituting σ with tan(σ), and numerically showed that this new relation is effective for larger σ. Setting $M_A$=1, i.e., $\upsilon = B/[3\times4\pi\rho]^{1/2}$, the improved CF relation gives $\sigma = 30°$, which is almost identical to the value we observed.

However, it is not only $M_A$ that affect the observed angle dispersion. Projection effects (while the line of sight is not perpendicular to the mean field) and non-turbulent structures of the B field also contribute to the angle dispersion. Non-turbulent B field structures include, e.g., those caused by stellar feedback (which causes the large angle dispersion in, for example, Perseus; Bally et al. 2008) and Galactic B field structures (note that the stars involved in the polarimetry data spread out over hundreds of pc along a line of sight). Recently, new analytical technologies have been developed to remove non-turbulent structures in the polarimetry data (Hildebrand et al. 2009; Houde et al. 2009). With all these non-turbulent factors that can significantly increase the B-field dispersion, we observed a B-field dispersion (*31°*) only nearly equivalent to the trans-Alfvénic condition assuming turbulence is the only force that can deviate field directions. This means that the turbulence should be sub-Alfvénic.

While our result, among other evidence shown in the Introduction, points toward the picture with sub-Alfvenic turbulence, there are observations which are used to support the scenario of super-Alfvénic clouds. However, these claims are based on assumptions which may not be fulfilled (see section 4.1). We will show that the relatively simple sub-Alfvénic model in Fig. 3 is possible to explain not only our observation but also the empirical cloud contraction density threshold (section 4.2) and the "hub-filament" cloud structures (section 4.3).

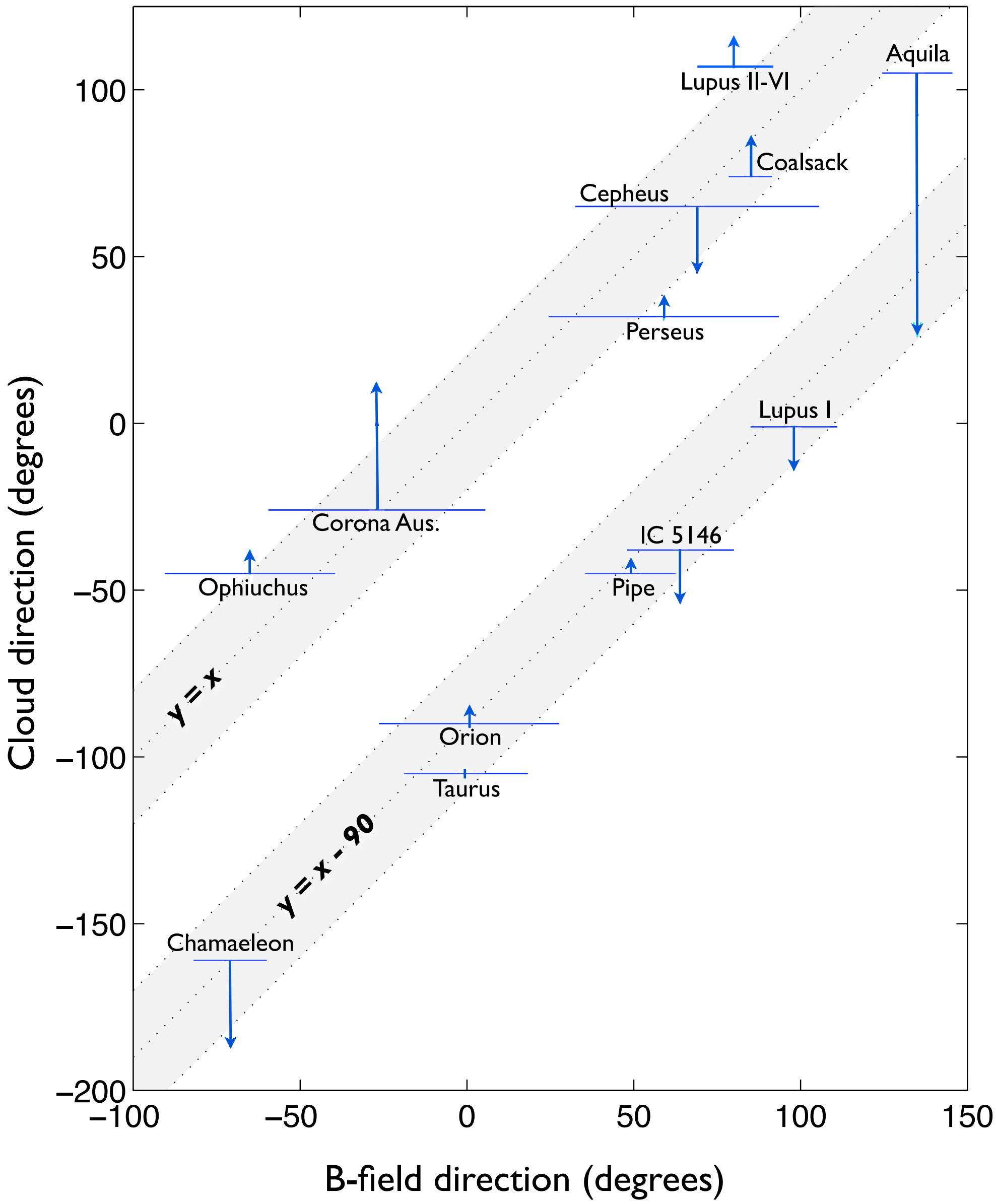

**Fig. 4 Cloud versus B-field directions.** The direction range of a cloud traced by the different autocorrelation contours is shown by the vertical arrow; the tail and head represent, respectively, the directions at low[2] and high[3] densities. The B-field direction IQRs are shown by the horizontal error bars. The width of each shaded zones shows the mean of the 13 IQRs, 42°.

---

[2] Traced by the blue contours in the middle column of Fig. 1 & 2.

[3] Traced by the red contours in the middle column of Fig. 1 & 2, except for Cepheus, where the red contour is roundish and the green contour is used instead.

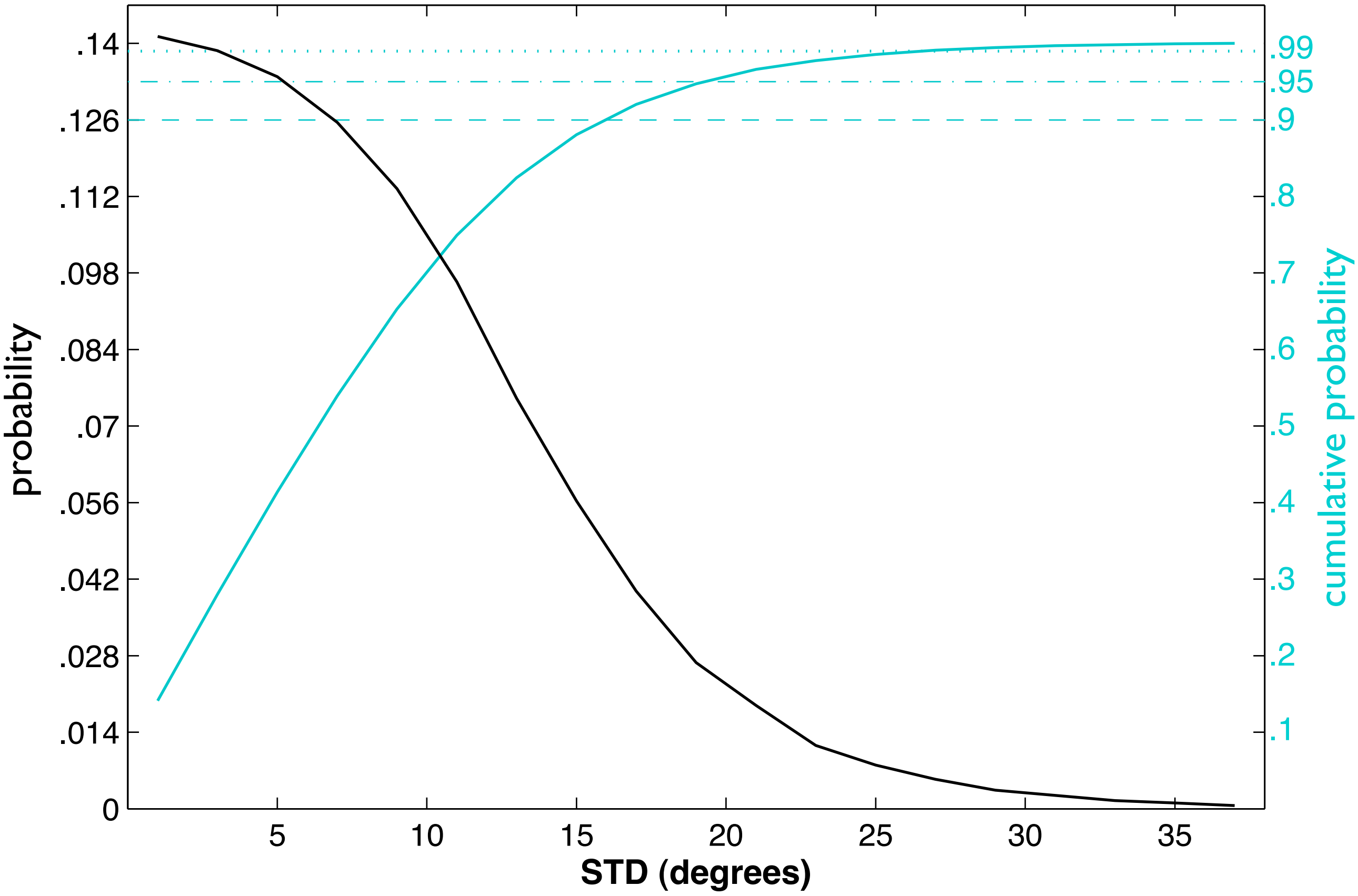

**Fig. 5 The probability of cloud-field alignment from Monte Carlo simulations and Bayesian analysis** (section 3). Assuming that cloud directions follow a double-gaussian distribution with a given standard deviation (STD) and peak at 0° and 90° from the B field, the darker curve shows the probabilities (left Y-axis) of various STD given the observed condition that all 13 cloud directions are within 30° from being either parallel or perpendicular to B-field directions. The cumulative probability (lighter curve; right Y-axis) is also shown. The dotted, dot-dash and dashed lines indicate respectively 0.99, 0.95 and 0.9 of the cumulative probability.

### 4.1 On the observational "evidence" of super-Alfvénic clouds

In this section we shortly discuss the main arguments which have been used so far as empirical evidence for molecular clouds being in a super-Alfvénic state.

*Power spectra indices of cloud column densities*

One analysis commonly used in support for super-Alfvénic clouds is the column density power spectra suggested by Padoan et al. (2004). Their simulations show that Alfvénic flows provide a power law index for column density of 2.25, while highly super-Alfvénic clouds have an index around 2.7. Comparing with these simulations, they concluded that Perseus, Taurus, and Rosetta are all super-Alfvénic, because the power-law indices of their $^{13}CO$ maps are around 2.75.

However, first, the power-law index depends on the exact set-up of the simulations. For example, cloud simulations of Collins et al. (2012) show much shallower spectra of column densities, and the power-law indices are *not* correlated with magnetic Mach numbers. Second, there is a spectrum of the observed power law indices between 2 and 3; see Table I of Schneider et al. (2011) for example. Third, different tracers can result in very different power-law indices. For example, the indices of Perseus, Taurus, and Rosetta are, respectively, 2.16, 2.20, and 2.55 when dust extinction is used to trace the column densities (Schneider et al. 2011). Schneider et al. (2011) concluded that the indices probed by dust extinction are usually significantly lower than those probed by CO. With careful comparison of extinction, thermal emission and CO maps of Perseus, Goodman et al. (2009) concluded that dust is a better tracer of column density than CO, because it has no problems of threshold density, opacity, and chemical depletion.

Furthermore, as we have seen in section 2.1, filamentary structures are equivalent to anisotropic autocorrelation functions (which is how we defined the filament directions) and, thus, anisotropic power spectra. The averaged index in this case depends on how the filamentary structure is projected on the sky. These considerations put in doubt the conclusion that empirical column density power laws support super-Alfvénic states of molecular clouds.

*Cloud-core field orientations*

Stephens et al. (2011) used the fact that cloud-core fields are not aligned with the Galactic disc to conclude that core B-fields must have decoupled from the Galactic B fields. This conclusion has been used as another indication against the strong-field scenario. Their argument relies on the assumption that Galactic fields are largely aligned with the disc plane, which, however, is not the case at the scale of cloud accumulation length as we will show in the following.

Figure 6 in Stephens et al. (2011) shows an angle distribution of almost all the polarimetry detections from the Heiles (2000) catalog, and the distribution clearly peaks at the direction of the Galactic disc plane. Note that this plot contains stars from distances of 140 pc to several kpc, and thus shows the (Stokes) mean B-fields from various scales because the polarization of a star samples the entire sight line (Fig. 6). As a result, one cannot establish from their plot whether the B-field coherence happens at every scale or only at some certain scale ranges. To distinguish between the two conditions, in Fig. 6, we plot similar polarization distributions but only for stars with distances within 100-pc bins centered at, respectively, 100, 300, 700, 1500, and 2500 pc in distance. We also use the optical data archive of Heiles (2000). We exclude data for which the ratio of the polarization level to its uncertainty is less than 2. The numbers of stars in each distance range are, from near to far, 1072, 339, 116, 82, and 51. At 100-pc scale the distribution is very flat, i.e., Galactic B-fields can have any direction. As shown in Fig. 6, the so-called "coherent Galactic B field" only appears at scales above 700 pc, where structures at smaller scales are averaged out. Also shown (with a dashed line) is the distribution of the B-field directions from 52 cloud cores at pc to sub-pc scales from Stephens et al. (2011). They concluded that the core B-fields must have decoupled from the Galactic B-fields, because the direction distribution of the core fields is not as peaked as their Fig. 6. However, as the accumulation length of even a GMC is only ~ 400 pc (Williams, Blitz & McKee 2000), the core B-fields and Galactic fields above 400-pc scale are irrelevant. In fact, the distributions of the core fields and the Galactic fields at 100-300 pc scales are very similar in Fig. 6. With the same archives, Li et al. (2009) have studied the core fields and the polarization within 100 - 200 pc (accumulation length) from each core, and showed a significant correlation (their Figure 2). This means that the *structures of Galactic B-fields* at the scale of cloud formation are preserved in the cores.

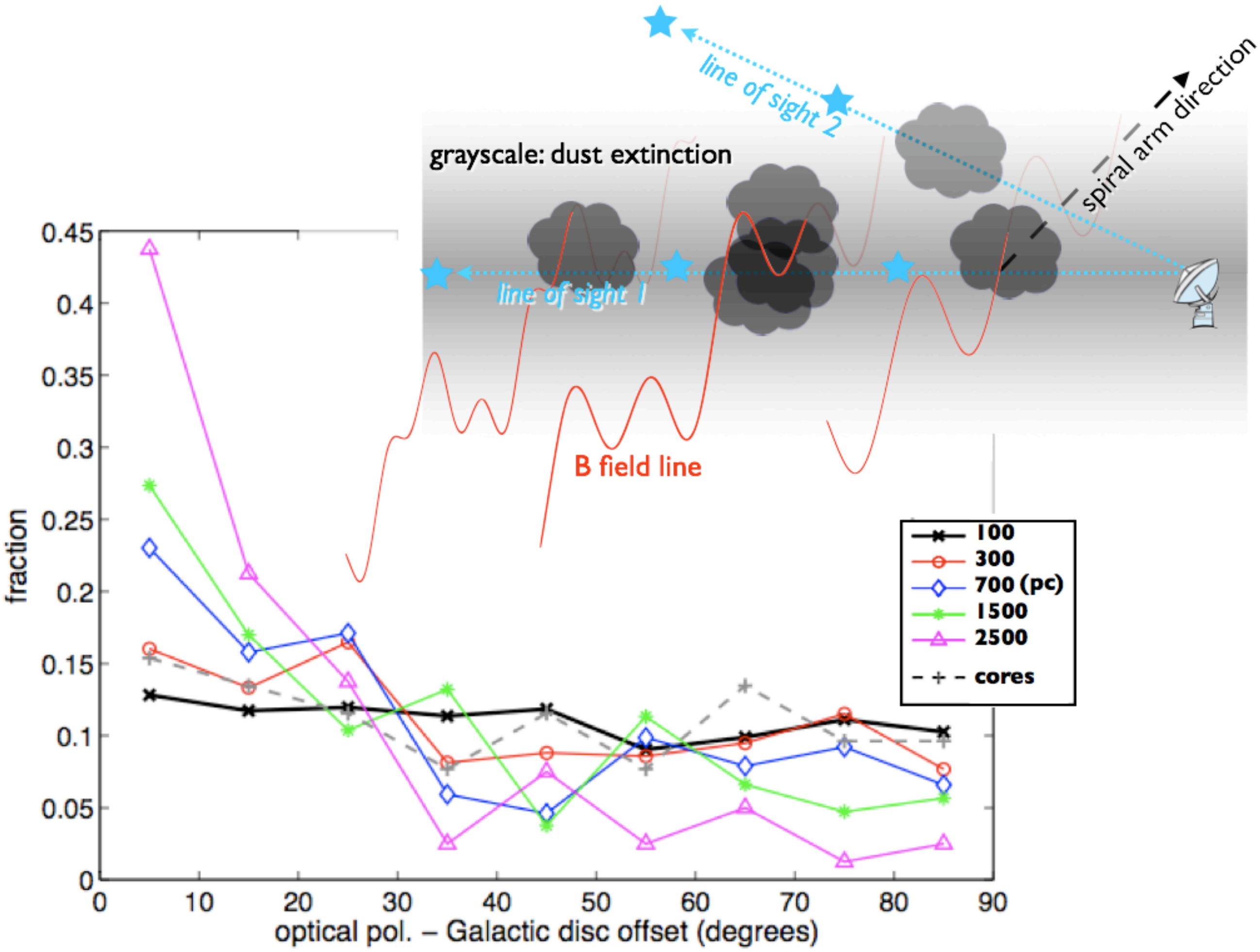

**Fig. 6** *Upper panel*: An illustration of the fact that galactic B-fields (red lines) follow spiral arms (dark dashed arrow) and anchor clouds (Li & Henning 2011; Li et al. 2009), but have rich structures perpendicular to the galactic disc, as shown in Fig. 1 & 2. Compared to line of sight 1, line of sight 2 goes through less galactic mass other than one particular cloud. *Stars with larger distance provide averaged field directions for larger scales*.

*Lower panel*: We sample the stars with reliable polarization detections from the 100-pc bins centered at distance 100, 300, 700, 1500, and 2500 pc. The plots show that the Galactic B-field is more coherent at larger scales (> 700 pc), but is almost random at the scale ~100 pc. Also plotted is the distribution of the B-fields from cloud cores (dashed line) at pc to sub-pc scales probed with thermal dust emission (Dotson et al. 2010). *The core field distribution is very similar to that of Galactic fields at 100-300 pc scales*, the size of the accumulation length of a typical molecular cloud.

### 4.2 Magnetic critical density, star formation threshold and cloud contraction threshold

Assuming that Fig. 3 presents the correct scenario for cloud formation, we want to investigate whether a magnetic critical density can explain the empirical star formation and cloud contraction thresholds.

The cloud formation model with B-field energy that dominates turbulent and self-gravity energies implies a constant field strength/morphology during cloud formation. The star-forming sites within a cloud apparently have to be massive enough to overcome this magnetic pressure in order to contract in all directions. If magnetic pressure is indeed the major force that regulates the gravitational contraction, we should expect the magnetic critical density to be close to the density threshold for star formation. We can test this idea by comparing the empirical star formation threshold (Lada et al. 2010; Heiderman et al. 2010) with the critical column density required to contract under the observed B-field strength.

Another indicator of gravitational contraction is the shape of the probability density function (PDF) of cloud column densities. Numerical simulations show that this PDF of non-gravitating clouds is log-normal for both sub- and super-Alfvénic clouds (e.g., Li et al. 2008; Collins et al. 2012). While self-gravity is also simulated, the PDF of high-density regions with dominant gravitational energy deviates from the log-normal function followed by the low-density PDF (e.g., Nordlund & Padoan 1999). This log-normal type PDF and the deviation are indeed observed (e.g., Kainulainen et al. 2009; Froebrich & Rowles 2010) and the transition point, the cloud contraction threshold, can also be compared with the magnetic critical density.

*Magnetic critical density*

In a recent review of molecular cloud magnetic field measurements, Crutcher (2012) summarized the Zeeman measurements from the past decade in a $B_{los}$ (line-of-sight field strength)-versus-$N_H$ (H column density) plot. We show this plot in Fig. 7 and highlight the column density ranges of the Gould Belt clouds (as in Fig. 1 & 2), cloud cores and ICM (traced by $H_I$ data). Based on a Bayesian analysis, Crutcher et al. (2010) concluded two most probable scenarios for ICM B-field strength: (1) constant strength around 10 μG and (2) any strength between 0 and 10 μG with a median of 6 μG. Most importantly, the B-field strength remains relatively constant in both cases over column densities from the ICM to the lower density regions in clouds. This is interpreted as evidence that *gas can only accumulate along the B fields* during cloud formation (Crutcher et al. 2010; Crutcher 2012), the same as proposed in Fig. 3.

For roughly $N_H > 10^{21}$ cm$^{-2}$ , the B-field strength increases with $N_H$, which implies that self-gravity is able to compress the field lines after accumulating adequate mass along the fields. The slanted solid line in Fig. 7, B (μG) = $3.8\times10^{-21}$ $N_H$ (cm$^{-2}$), is a theoretical calculation of the balanced condition between magnetic pressure and self-gravity, the so-called *magnetic critical condition* (Crutcher 2012; Nakano & Nakamura 1978). Since the cloud mass is accumulated along the B fields, the cloud shape should be more sheet-like (e.g., Shetty & Ostriker 2006) instead of spherical, so statistically the observed column density should be twice the value for calculating the criticality (the column density observed with a sight line aligned with the B field) due to projection effects (Shu et al. 1999). Taking the projection effects into account, we add to Fig. 7 the corrected magnetic critical condition,

$$B = 1.9\times10^{-21}\ N_H \qquad \text{(I)}$$

Assuming an equipartition between turbulent and magnetic energies, which employs the upper limit of turbulent energy of sub-Alfvénic clouds, and magnetic virial equilibrium, $2T + M + U = 0$ (where $T$, $M$ and $U$ are, respectively, kinetic, magnetic and gravitational potential energies; McKee et al. 1993), the *critical condition* becomes

$$B = 1.1\times10^{-21}\ N_H \qquad \text{(II)}$$

Equations (I) and (II) give the lower and upper limits of the critical column density. For B = 10 μG, the critical column density ranges between $N_H$ =[5.3 9.1]$\times10^{21}$ cm$^{-2}$; for B = 6 μG, the range is $N_H$ =[3.2 5.5]$\times10^{21}$ cm$^{-2}$ (Fig. 7).

*Star formation threshold*

Recently, Lada et al. (2010) surveyed the number of young stellar objects (YSOs) in nearby molecular clouds and showed that it correlates *better* with the mass of *dense gas* than with the total gas mass in the clouds. The best correlation occurs when the dense gas is defined with extinction above $A_v$ = 7.3±1.8 mag. This led Lada et al. to suggest that a threshold for star formation exists. We should convert Av = 7.3 mag to column density, $N_H$, so it can be compared with the critical column density. Assuming the standard ratio $N_H/A_v = 1.87 \times 10^{21}$ cm$^{-2}$ (Vrba & Rydgren 1984; Bohlin et al. 1978), $A_v$ = 7.3±1.8 mag can be converted to $N_H = (1.35 \pm 0.35) \times 10^{22}$ cm$^{-2}$ (Fig. 7).

Another survey of the star formation threshold comes from Heiderman et al. (2010). They define the star formation threshold to be 8.1±0.9 mag by the density where the observed SFR (star formation rate)-column density relation changes from a power-law (low density) to a linear relation (high density). Their estimate can be converted to $N_H = (1.55 \pm 0.15) \times 10^{22}$ cm$^{-2}$ (Fig. 7). Observations of individual Gould belt clouds have suggested similar thresholds for core formation (e.g., André et al. 2010, Johnstone et al. 2004, Onishi et al. 1998)

*Cloud contraction threshold*

For most molecular clouds, column density PDFs can be well-fitted by log-normal functions at low column densities, but power-law-like wings are common at higher column densities. Based on the column densities traced by near-IR extinction data, Kainulainen et al. (2009) and Froebrich & Rowles (2010) observed that this transition occurs around $A_v$ = 2-5 mag and $A_v$ = 4.5-7.5 mag, respectively. This $A_v$ value can be converted to $N_H = [3.7\ 9.4] \times 10^{21}$ cm$^{-2}$ and $N_H = [8.4\ 14.0] \times 10^{21}$ cm$^{-2}$ (Fig. 7).

Fig. 7 shows that the star formation threshold is about 1.5 times larger than the cloud contraction threshold. While the uncertainties in the $A_v$ measurements might contribute to this difference, it also seems quite reasonable because the cloud contraction threshold traces the onset of cloud contraction while star formation must happen afterwards when the clouds become denser. We also note that Krumholz et al. (2012) pointed out that the better correlation between YSO numbers and cloud mass above $A_v$ = 7.3 mag (Lada et al. 2010) does not mean that stars cannot form below $A_v$ = 7.3 mag. In fact, 2/3 of the YSOs in the study of Lada et al. are found with the densities below this value (Krumholz et al. 2012). Lada et al. (2010) argue that, on the other hand, stars might have formed in regions with higher densities and then migrated away. We conclude that the cloud contraction threshold is probably a better tracer for the minimum column density needed to overcome magnetic and turbulent pressures and is indeed comparable to the range of critical column density we derived (Fig. 7). Therefore, B fields may play an important role in setting the threshold for cloud contraction observed in PDFs of molecular clouds.

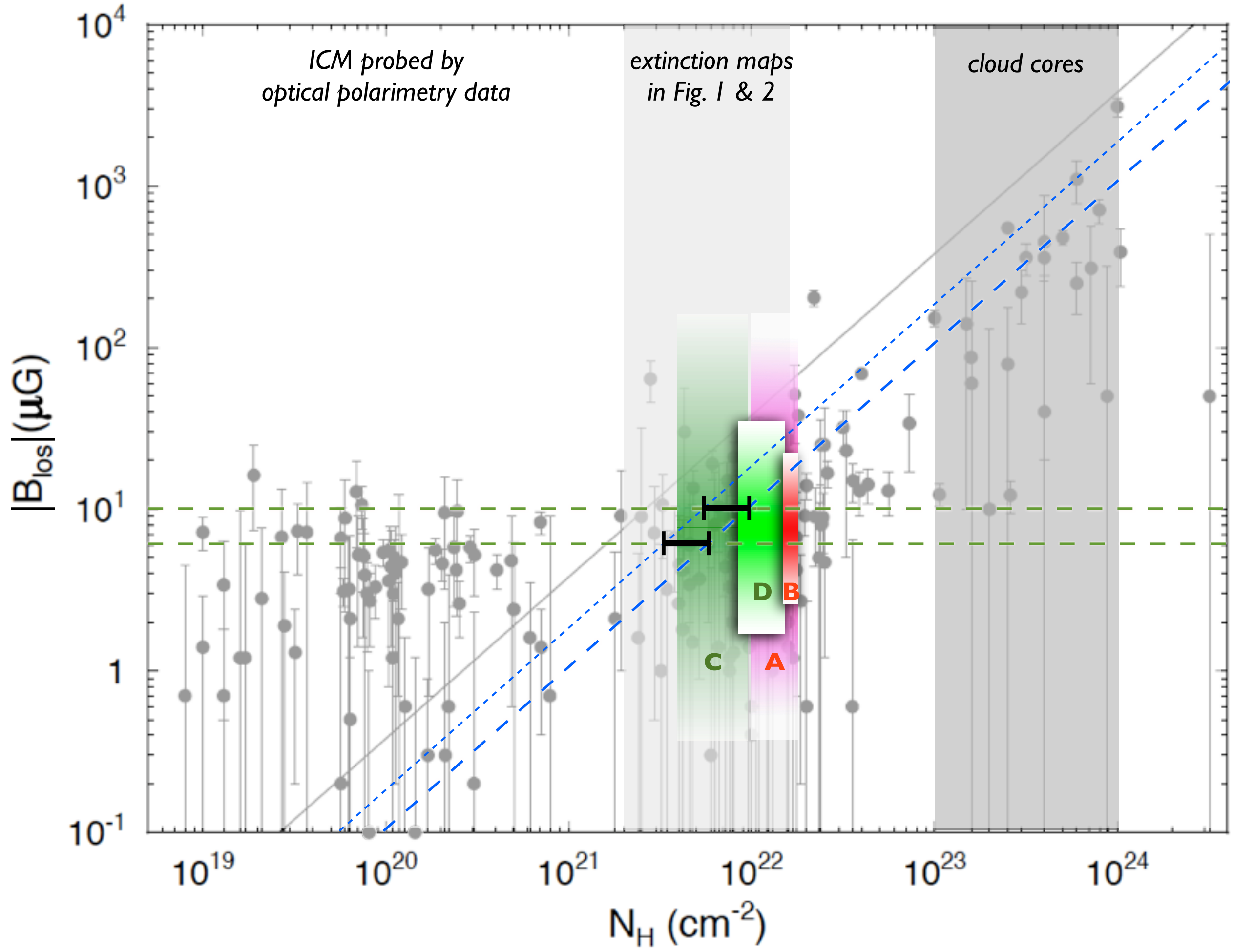

**Fig. 7 The agreement between the magnetic critical density and gravitational contraction threshold.** On top of the plot of line-of-sight field strength ($B_{los}$) versus column density ($N_H$) from Crutcher (2012), we note the three column density zones that are related to the inter molecular cloud media (ICM), cloud cores (dark shaded zone), and Gould Belt clouds as probed by near-infrared extinction maps (Fig. 1 & 2)., The field strength is quite constant from the ICM to lower-density regions of the clouds (Crutcher et al. 2010). The two horizontal lines mark 10 and 6 μG respectively (see section 4.2).

The slanted solid line is the theoretical magnetic critical condition from Crutcher (2012). Applying a projection-effect correction (Shu et al. 1999) to it, we obtain the dotted line. Assuming an equipartition condition between magnetic and turbulent energies, we obtain the upper limit of the critical condition (slanted dashed line). The two "H" shapped symbols mark the range of possible critical densities for, respectively, B = 10 and 6 μG.

The emprical star formation threshold, $A_V$ = 7.3±1.8 mag from Lada et al. (2010) and $A_V$ = 8.1±0.9 mag from Heiderman et al. (2010) are shown by the zones labled A and B. Zone C (Kainulainen et al. 2009) and D (Froebrich & Rowles 2010) show where the observed column density PDF turns from log-normal to power-law like.

### 4.3 The "hub-filament" cloud structures

Our sample covers all the nearby clouds examined by Myers (2009), who concluded that the clouds can be often described by a "hub-filament" morphology. The clouds have high-density elongated "hubs", which host most of star formation in the clouds, and lower-density "parallel filaments" direct *mostly* along the short axes of the hubs. Myers (2009) suggested that the parallel filaments are due to layer fragmentation. The main result of our work, i.e., that elongated/filamentary structures tend to be aligned either parallel or perpendicular to the ambient ICM B fields, suggests that the gas layers in this scenario must host ordered and dynamically dominant B fields, as shown by Nagai et al. (1998).

Note that the two types of B field-regulated filaments in Fig. 3 can also explain the hub-filament structures. The two types of filaments may form at the same time, with the denser and more massive filaments (hubs) perpendicular to the B field and finer filaments in the vicinity aligned with the field (see observations from e.g., Palmeirim et al. 2013 and Goldsmith et al. 2008 and simulations from Nakamura & Li 2008 and Price & Bate 2008), a picture identical to the hub-filament structure described by Myers (2009). In this scenario, hubs and filaments should always be perpendicular to each other, but their sky projections are not necessarily so. This means that at least some of the exceptions, i.e., non-perpendicularity (e.g., Figure 9 of Myers 2009), can be explained by projection effects. For the same reason, projections will also affect the dispersion of the B field-cloud alignment shown in Fig. 4, and this is the reason we have carried out Monte Carlo simulations to study the 3-D alignment (Fig. 5). The hub-filament system can be an alternative explanation (besides rotation during contraction; see section 2.1) of the larger differences between cloud directions defined by different column densities (e.g., see Corona Australis and Aquila in Fig. 2).

The hub-filament could be a self-similar structure. For example, the Herschel telescope resolved part of the Pipe nebula (Fig. 1) with 0.5′ resolution and showed that the hub fragments formed into a network of perpendicular filaments (Peretto et al. 2012). The network is aligned with the mean B fields' direction shown in Fig. 1. A similar analysis as we performed here for the Herschel Gould Belt data will be of interest. Tassis et al. (2009) surveyed 32 cloud cores with 20″ resolution (Caltech Submillimeter Observatory), which showed that the elongated cores tend be perpendicular to the B fields. Li et al. (2011) observed one of these core regions (NGC 2024) with 3″ resolution (Submillimeter Array) and found filaments perpendicular to the core (i.e., a hub-filament structure). A survey of core vicinities with high angular resolution and sensitivity, as performed by Li et al. (2011), is necessary to tell whether NGC 2024 is a special case.

**5. Summary**

Filamentary structures are ubiquitous in molecular clouds and various mechanisms have been proposed for their formation. Inspired by the fact that cloud B fields are ordered and aligned with the ambient ICM fields (Li & Henning 2011; Li et al. 2009; Han & Zhang 2007) and the fact that B-field strengths are quite constant from ICM to clouds (Crutcher 2012; Crutcher et al. 2010), we studied the alignments between filamentary molecular clouds and local ICM B fields and reached the following conclusions:

1. Filamentary clouds in the Gould Belt have bimodal orientations with respect to local ICM B fields: they are close to be either parallel or perpendicular to each other, with a typical offset less than 20° (95 % confidential range; section 3). This indicates that the ICM B fields are strong enough to guide gravitational contraction to form flat condensations *perpendicular* to them and strong enough to channel turbulence to result in filaments *aligned with* them.
2. In the scenario of filament formation with dynamically important B fields (Fig. 3), the field strength should not change significantly during the cloud formation process, which is consistent with the constant ICM field strength estimated by the Zeeman measurements (Fig. 7). This field strength sets a critical column density, $N_H$ =[3.2 9.1]×$10^{21}$ $cm^{-2}$, that is comparable with the empirical threshold of gravitational contraction, $N_H$ =[3.7 14]×$10^{21}$ $cm^{-2}$ (Kainulainen et al. 2009; Froebrich & Rowles 2010), and is a little smaller than the empirical star formation threshold, $N_H$ = (1.35±0.35)×$10^{22}$ $cm^{-2}$ (Lada et al. 2010; Heiderman et al. 2010).

3. We show that dynamically important B-fields can give rise to the typical hub-filament cloud morphology observed by Myers (2009).
4. The hypothesis of a coherent Galactic B field aligned with the disc plane is realistic only for the mean field at scales > 700 pc. At smaller scales, the Galactic B field can have any orientation (see Table I and Fig. 6). So even though the orientations of cloud filaments (this work) and cloud B fields (Li et al. 2009) both significantly correlate with the local Galactic (ICM) B fields, they are not necessarily correlated with the Galactic disc plane.

**Acknowledgments**

This work was supported by the Deutsche Forschungsgemeinschaft priority program 1573 ("Physics of the Interstellar Medium"), project #46, The Anisotropy in Interstellar Media. We thank the referee for a very constructive report. We appreciate the valuable comments from Bruce Elmegreen, Zhi-Yun Li and Frank Shu.